
\input phyzzx

%
\catcode`\@=11 
\def\papers{\papersize\headline=\paperheadline\footline=\paperfootline}
\def\papersize{\hsize=40pc \vsize=53pc \hoffset=0pc \voffset=1pc
   \advance\hoffset by\HOFFSET \advance\voffset by\VOFFSET
   \pagebottomfiller=0pc
   \skip\footins=\bigskipamount \normalspace }
\catcode`\@=12 
\papers

\def\to{\rightarrow}
\def\half{\textstyle{1\over 2}}

\vsize=21.5cm
\hsize=15.cm

\tolerance=500000
\overfullrule=0pt

\pubnum={PUPT-1411 \cr
hep-th/9307052 \cr
July 1993}

\date={}
\pubtype={}
\titlepage
\title{ASPECTS OF EXACTLY SOLVABLE QUANTUM-CORRECTED
2D DILATON-GRAVITY THEORIES}
\author{{
Adel~Bilal}\foot{
 on leave of absence from
Laboratoire de Physique Th\'eorique de l'Ecole
Normale Sup\'erieure, \nextline 24 rue Lhomond, 75231
Paris Cedex 05, France
(unit\'e propre du CNRS)\nextline
e-mail: bilal@puhep1.princeton.edu
}}
\address{\it Joseph Henry Laboratories\break
Princeton University\break
Princeton, NJ
08544, USA}

\vskip 3.mm
\abstract{After reviewing the basic aspects of the exactly
solvable quantum-corrected dilaton-gravity theories in two
dimensions, we discuss a (subjective) selection of other
aspects: a) supersymmetric extensions, b) canonical
formalism, ADM-mass, and the functional integral measure,
and c) a positive energy theorem and its application to the
ADM- and Bondi-masses.}

\vskip 1.cm

\centerline{\it Based on Talks given at }
\centerline{\it ``Strings 93", Berkeley, May 1993}
\centerline{\it and at the Santa Barbara conference on }
\centerline{\it ``Quantum Aspects of Black Holes", June 1993
}

\endpage
\pagenumber=1

 \def\PL #1 #2 #3 {Phys.~Lett.~{\bf #1} (#2) #3}
 \def\NP #1 #2 #3 {Nucl.~Phys.~{\bf #1} (#2) #3}
 \def\PR #1 #2 #3 {Phys.~Rev.~{\bf #1} (#2) #3}
 \def\PRL #1 #2 #3 {Phys.~Rev.~Lett.~{\bf #1} (#2) #3}
 \def\CMP #1 #2 #3 {Comm.~Math.~Phys.~{\bf #1} (#2) #3}
 \def\IJMP #1 #2 #3 {Int.~J.~Mod.~Phys.~{\bf #1} (#2) #3}
 \def\JETP #1 #2 #3 {Sov.~Phys.~JETP.~{\bf #1} (#2) #3}
 \def\PRS #1 #2 #3 {Proc.~Roy.~Soc.~{\bf #1} (#2) #3}
 \def\IM #1 #2 #3 {Inv.~Math.~{\bf #1} (#2) #3}
 \def\JFA #1 #2 #3 {J.~Funkt.~Anal.~{\bf #1} (#2) #3}
 \def\LMP #1 #2 #3 {Lett.~Math.~Phys.~{\bf #1} (#2) #3}
 \def\IJMP #1 #2 #3 {Int.~J.~Mod.~Phys.~{\bf #1} (#2) #3}
 \def\FAA #1 #2 #3 {Funct.~Anal.~Appl.~{\bf #1} (#2) #3}
 \def\AP #1 #2 #3 {Ann.~Phys.~{\bf #1} (#2) #3}
 \def\MPL #1 #2 #3 {Mod.~Phys.~Lett.~{\bf #1} (#2) #3}

\def\d{\partial}
\def\dpl{\partial_+}
\def\dm{\partial_-}
\def\dpm{\partial_\pm}
\def\Scl{S_{\rm cl}}
\def\Sm{S_{\rm matter}}
\def\Sanom{S_{\rm anom}}

\def\f{\phi}
\def\r{\rho}
\def\o{\omega}
\def\O{\Omega}
\def\x{\chi}
\def\s{\sigma}
\def\sp{\sigma^+}
\def\sps{\sigma^+_s}

\def\spo{\sigma^+_0}

\def\sm{\sigma^-}
\def\spm{\sigma^\pm}
\def\sms{\sigma^-_s}

\def\l{\lambda}
\def\t{\tau}
\def\ix{\int {\rm d}^2 x}
\def\ixt{\int {\rm d}^2 x {\rm d}^2 \theta}
\def\is{\int {\rm d}^2 \sigma}
\def\rg{\sqrt{-g}}
\def\k{\kappa}

\def\eoms{equations of motion\ }
\def\m{\mu}
\def\n{\nu}
\def\tmn{T_{\mu\nu}}
\def\N{\nabla}
\def\zt{{\tilde Z}}
\def\S{\Sigma}
\def\e{\epsilon}
\def\eb{\bar\epsilon}
\def\a{\alpha}
\def\b{\beta}
\def\g{\gamma}
\def\gf{\gamma_5}
\def\G{\Gamma}
\def\NS{ {/\kern -0.60em \nabla } }
\def\M{{\cal M}}
\def\Or{{\cal O}}
\def\kk{{\k\over 4}}
\def\mb{M_{\rm B}}
\def\stp{\s\to\infty}

\def\sep{\s =+\infty}
\def\sem{\s =-\infty}
\def\pf{\Pi_\f}
\def\pr{\Pi_\r}
\def\pz{\Pi_Z}
\def\ca{{\cal C}_A}
\def\cb{{\cal C}_B}

{ \chapter{Basic Aspects}}

\section{Introduction}

Our starting point is the classical action for dilaton
gravity in two dimensions as written by Callan, Giddings,
Harvey and Strominger (CGHS)
\REF\CGHS{C. Callan, S. Giddings, J. Harvey and A.
Strominger, \PR D45 1992 R1005 .} [\CGHS]
$$
\Scl = {1\over 2\pi} \ix \rg \left[ e^{-2\f}\left(
R+4(\nabla\f)^2 + 4\l^2\right) -\half \sum_{i=1}^N
(\nabla f_i)^2\right]\ .
\eqn\ui$$
Here $\f$ is the dilaton field with $ G\equiv e^{\f}$
playing the role of the gravitational coupling constant,
$\l^2$ is referred to as cosmological constant and the
$f_i$ are $N$ massless conformally coupled matter fields.
This action
admits classical (non-radiating) static black hole solutions
$$ \eqalign{
{\rm d}s^2 &= -{{\rm d}x^+ {\rm d}x^-\over
{m\over \l} -\l^2 x^+ x^-} = {-{\rm d}\t^2 +{\rm d}\s^2
\over 1+{m\over \l} e^{-2\l\s}} \cr
e^{-2\f}&={m\over \l} -\l^2 x^+ x^- = {m\over \l} +
e^{2\l\s}\ .}
\eqn\uii$$
The $x^\pm$ are Kruskal type coordinates, related to the
Schwarzschild type coordinates $\s, \t$ by
$\l x^\pm=\pm e^{\l\pm \spm},\ \spm=\t\pm\s$. The metric in
the latter coordinates is asymptotically Minkowskian as
$\s\to\infty$. The parameter $m$ is the black hole mass, and
the $m=0$ solution where $\f=-\l\s$ is called the linear
dilaton vacuum (LDV).

The goal then is to quantize the theory described by this
action $\Scl$. If the number of matter fields is
different from 24, $N\ne 24$, one has to include various
contributions to the conformal anomaly. Thus we add
\foot{A possible term $\mu^2\int \rg$ is supposed to be
fine-tuned to vanish.}
$$\Sanom = -{\k\over 8\pi} \ix \rg R{1\over \nabla^2} R
\eqn\uiii$$
We will keep $\k$ as a
parameter to be determined later on. Note that $\Sanom$ is
${\cal O}(e^{2\f})\equiv {\cal O}(G^2)$ with respect to the
gravitational part of $\Scl$ and may be thought of as the
one-loop contribution of the matter fields. We will refer to
$\Scl + \Sanom$ as $S_{\rm CGHS}$ (with  $\k_{\rm
CGHS}={N\over 12}$).

\section{Conformal invariance and transformation to free
fields}

\subsection{Conformal gauge}

Let us first choose conformal gauge, $g_{++}=g_{--}=0,\,
g_{+-}=-\half e^{2\r}$. Then
$$\eqalign{
\Scl&={1\over \pi}\is \left[ e^{-2\f}\left(
2\dpl\dm\r-4\dpl\f\dm\f+\l^2e^{2\r}\right) +{1\over 2}
\sum_{i=1}^N \dpl f_i \dm f_i\right]\ , \cr
\Sanom&= -{\k\over \pi} \is \dpl \r\dm\r\ .}
\eqn\uiv$$
The \eoms derived in conformal gauge must be supplemented
by the $g_{++}$ and $g_{--}$ \eoms as constraints:
$$\eqalign{
T_{++}=&T_{--}=0\ ,\cr
T_{\pm\pm}=&e^{-2\f}\left( 4\dpm\f\dpm\r-2\dpm^2\f\right)
-\k\left( \dpm\r\dpm\r-\dpm^2\r\right) \cr
&+{1\over 2} \sum_{i=1}^N \left( \dpm f_i\right)^2\ .}
\eqn\uv$$
Note that for $\k>0$ the kinetic term of $\Scl +\Sanom$ is
degenerate, %
$$
\det\pmatrix{-4e^{-2\f} & 2e^{-2\f}\cr
2e^{-2\f}& -\k} = 4 e^{-2\f} \left( \k - e^{-2\f} \right) =
0 \quad {\rm at} \quad e^{-2\f} =e^{-2\f_c}\equiv \k \ .
\eqn\uvi$$
We expect something singular to happen when
$\f=\f_c$.

\subsection{Conformal invariance}

Since we are dealing with a theory of
gravity, we started with a diffeomorphism
invariant theory. Then we fixed conformal
gauge, leaving as symmetries the subgroup
of conformal diffeomorphisms $\sp\to
f^+(\sp),\ \sm\to f^-(\sm)$. Quantization should preserve
these conformal symmetries. In particular, we need to ensure
that the resulting theory is a $c_{\rm tot}=0$
conformal theory. The latter is a
necessary condition that relies only on the short distance
properties of the quantum theory. They may be
inferred even though the full quantum theory might not be
known. For $\k>0$, a non-trivial complication is the presence
of the critical value of $\f$ where we expect a singularity.
Typically $\f=\f_c$ on some line. Although the presence of
this boundary type line complicates the elaboration of a
complete quantum theory, it should not affect the
short-distance singularities of the propagators away from
it. Hence we should be able to check whether or not a
theory is conformally invariant away from this line. This
is the approach taken here (see
\REF\BC{A. Bilal and C. Callan, \NP B394 1993 73 .}
\REF\ALW{S. de Alwis,
 \PL B289 1992 278 ,
 \PL B300 1993 330 .}
refs. [\BC] and [\ALW]): we will display a class of
theories that are conformally invariant, at least when we
need not consider the line of singularity, or if it is
absent as for $\k<0$.

\subsection{Transformation to free fields}

The kinetic part of $\Scl+\Sanom$ can be written as
$$
S_{\rm kin}={1\over \pi}\is  e^{-2\f}\left(
-4\dpl\f\dm\f+2\dpl\r\dm\f+2\dpl\f\dm\r-\k\dpl\r\dm\r
\right) +\Sm\ .
\eqn\uvii$$
Here, we assume $\k>0$, while things work similar for
$\k<0$. Let now [\BC]\foot{Here we rescale $\x$ and $\O$
by a factor 2 with respect to ref. [\BC], and $\O$ also is
shifted by a constant.}
$$\o=e^{-\f}/\sqrt{\k}\ ,\quad \x
=\r+\o^2\ .
\eqn\uviii$$
Then
$$
S_{\rm kin}={1\over \pi}\is  \left[ -\k\dpl\x\dm\x+4\k
(\o^2-1)\dpl\o\dm\o \right]
 +\Sm
\eqn\uix$$
is diagonalized. We can bring the $\o$-kinetic term into a
standard form by a further (local) field redefinition:
$$\O=\o\sqrt{\o^2-1}-\log\left(
\o+\sqrt{\o^2-1}\right) +\half\left( 1-\log{\k\over
4}\right) \ \Rightarrow\ \d\O=2\sqrt{\o^2-1}\d\o \ ,
\eqn\ux$$
so that finally
$$
S_{\rm kin}={1\over \pi}\is  \left[ -\k\dpl\x\dm\x+\k
\dpl\O\dm\O +{1\over 2}\sum_{i=1}^N \dpl f_i \dm f_i
\right] \ .
\eqn\uxi$$
Note that $\x$ and $\O$ have opposite signature, but it now
seems that the kinetic term can no longer become singular.
What has happened to the singularity (for $\k>0$) at
$\f=\f_c$? Of course, it has been hidden in the
transformation from $\f$ to $\O$. This transformation
is not one to one (for $\k>0$) and we have a singularity
when ${\rm d}\O/{\rm d}\f =0$ which precisely happens at
$\f=\f_c$.

Not only the kinetic part of the action is very simple when
written in terms of the new fields $\x$ and $\O$, but also
the stress tensor:
$$
T_{\pm\pm}=-\k(\dpm\x)^2 + \k \dpm^2\x +\k(\dpm\O)^2
+{1\over 2} \sum_{i=1}^N \left( \dpm f_i\right)^2\ .
\eqn\uxii$$
These forms of the kinetic part of the action and of the
stress tensor are valid for both signs of $\k$, only the
precise form of the field transformations are different.

For $\k<0$ quantization is straightforward. For $\k>0$, we
disregard the subtleties connected with the presence of a
singular line $\f=\f_c$ for the moment and proceed with a
``naive" quantization. The kinetic part of the action then
shows that $\x$ and $\O$ have standard massless free field
propagators, and it is straightforward to compute the short
distance expansion of the stress tensor with itself. We
find that it generates a (continuum) Virasoro algebra with
central charge
$$c=1+1-12\k+N-26=N-24-12\k
\eqn\uxiii$$
which vanishes precisely if
$$\k={N-24\over 12}\ .
\eqn\uxiv$$
The field $\O$ contributes 1 to the central charge. $\x$
contributes $1-12\k$ since the $\x$ part of $T_{\pm\pm}$
has the Feigin-Fuchs form with background charge $\sim
\sqrt{\k}$. Furthermore, the matter fields just contribute
the usual $N$, while, although not written explicitly, we
also have ghost from the conformal gauge fixing, and they
contribute $-26$, as always.

Thus, with $\k=(N-24)/12$ we expect to have a conformal
field theory with vanishing central charge. One might worry
however, that when we transformed our fields from $\f$ and
$\r$ to $\x$ and $\O$, a complicated Jacobian would appear
in the functional integral, turning $\x$ and $\O$ into
interacting fields. This is not so , as we shall discuss
below. Indeed, the initial ``measure" for $\f$ and $\r$ is
precisely such that together with this Jacobian one obtains
a  standard free field measure for $\x$ and $\O$.

\section{$(1,1)$-operators and exact solutions}

So far we only discussed the kinetic part of the action.
There is also the ``interaction" part of the action which
contains the cosmological constant $\sim
\l^2 e^{-2\f}e^{2\r}$. This term behaves like a perturbation
of our conformal theory. If it is a marginal operator,
however, it will preserve the conformal invariance. A
necessary condition for a marginal operator is that it has
conformal dimension $(1,1)$. The cosmological constant
operator has indeed dimension $(1,1)$ classically, i.e. if
we do a Poisson bracket computation with $T_{\pm\pm}$, but
this is no longer true at the quantum level. It is easy to
see that the only operators (with no derivatives) of definite
conformal dimension $(\Delta,\Delta)$ are $\l^{2\Delta}
:e^{\a\x+\b\O}:$ with %
$$\Delta={\a\over 2}+{\a^2-\b^2\over \k}\ .
\eqn\uxv$$
Any of these operators with $\Delta=1$ (if truely
marginal) will lead to a conformal theory. However, in the
weak coupling limit, $e^{2\f}\to 0$, we want to recover the
classical dilaton-gravity action $\Scl$. Hence, we must
take $\a=2$ and $\b=-2$. It is easy to see that this
operator is indeed marginal. Then the full action reads:
$$
S={1\over \pi}\is  \left[ -\k\dpl\x\dm\x+\k
\dpl\O\dm\O + \l^2 e^{2(\x-\O)} +{1\over 2}\sum_{i=1}^N \dpl
f_i \dm f_i \right] \ .
\eqn\uxvi$$
Note that this action differs from the classical dilaton
gravity action only by higher order corrections, i.e. terms
that are ${\cal O}(G^2)={\cal O}(e^{2\f})$ with respect to
$\Scl$.

The \eoms that follow from the action \uxvi\ are very simple.
First of all, the $N$ matter fields are just free fields,
$\dpl\dm f_i=0$ with solution $f_i=f_i^+(\sp)+f_i^-(\sm)$.
For $\x$ and $\O$ we have:
$$\eqalign{
\dpl\dm (\x-\O)&=0\ ,\cr
\dpl\dm (\x+\O)&=-{2\over \k}\l^2 e^{2(\x-\O)}\ .}
\eqn\uxvii$$
The general solution reads
$$\eqalign{
2 (\x-\O)&=f^+(\sp)+f^-(\sm)\equiv \log \dpl \a^+(\sp)
+\log \dm\a^-(\sm)\ ,\cr
{\k\over 2}(\x+\O)&=-\l^2 \a^+(\sp)\a^-(\sm)+\b^+(\sp)
+\b^-(\sm)\ .\cr}
\eqn\uxviii$$
This solution is just as simple as the one for the
classical dilaton-gravity (i.e. the solution of the \eoms
derived from $\Scl$ alone) which reads $2(\r-\f)=\log \dpl
\a^+(\sp) +\log \dm\a^-(\sm)$ and $e^{-2\f}= -\l^2
\a^+(\sp)\a^-(\sm)+\b^+(\sp) +\b^-(\sm)$. Let us also give
the stress tensor when evaluated on the solutions \uxviii:
$$T_{\pm\pm}=-\dpm f^\pm \dpm \b^\pm +\dpm^2\b^\pm
+{\k\over 4} \dpm^2 f^\pm\ ,
\eqn\uxix$$
where $f^\pm \equiv \log \dpm \a^\pm$.

\section{Variations over the theme}

There are a few variations of the preceeding formalism
which we shall briefly discuss.

\subsection{Strominger's ``decoupled ghost" theory}

Strominger proposed
\REF\STRO{A. Strominger, \PR D46 1992 4396 .}
[\STRO] to define the measures for the different fields in a
functional integral with different metrics. The metric we
used so far is $g_{ij}=e^{2\r}\delta_{ij}$. It should be
used to define the measure for the matter fields $f_i$.
To define the measures for the dilaton field $\f$, the
conformal factor of the metric $\r$ and the
reparametrization ghosts, Strominger proposes to take a
different, Weyl rescaled metric
$g_{ij}=e^{-2\f}e^{2\r}\delta_{ij}$. Then the anomlaly
action for the latter fields will be constructed out of
$\r-\f$ instead of $\r$, and we have
$$\Sanom=-{1\over \pi} \is \left[ -{N\over 12} \dpl \r\dm\r
+2 \dpl (\r-\f)\dm(\r-\f) \right] \ .
\eqn\uxx$$
Consider first the case $N=0$, i.e. no matter fields. Then
the \eoms derived from $\Scl+\Sanom$ differ from those
derived from $\Scl$ only by terms $\sim \dpl\dm (\r-\f)$,
which vanishes by these \eoms : $\Sanom$ has no effect
on the solutions of the \eoms. Thus it was hoped that with
this modified anomaly action the ghosts, the dilaton and
$\r$-field would not contribute to the Hawking radiation.
When $N\ne 0$ however, $\dpl\dm (\r-\f)$ does no longer
vanish by the \eoms, and there seems to be no {\it a priori}
reason why the modified anomaly action should be more
physical than the original one (see the discussion in ref.
[\BC]).

Here we would like to show that the kinetic part of
Strominger's action can again be written in free field
form, and that after an appropriate improvement of the
cosmological constant term, one again obtains a conformal
theory. Indeed redefine fields as
$$\eqalign{
\o=e^{-\f}/\sqrt{\k}\ ,\ \ \x=\r+\o^2-2\log \o\ ,\cr
(\d\O)^2=4\left( \o^2-{\k+2\over \k} +{\k+2\over \k^2}
{1\over \o^2}\right) (\d\o)^2\ ,}
\eqn\uxxi$$
($\k=(N-24)/12$). Then the kinetic part of the action and
the stress tensor, when expressed in terms of the new
fields $\x$ and $\O$, take exactly the same form \uxi\ and
\uxii\ as before. Going through the same arguments about
how to modify the cosmological constant operator to turn it
into an exactly marginal operator, one finally arrives at
the same conformally invariant action \uxvi.
The reason that one obtains the same final action is very
simple.
Since Strominger's  action, just as the CGHS action, or the
actions to be discussed next, differ from the classical
dilaton-gravity action only by higher order corrections,
i.e. terms that are ${\cal O}(G^2)={\cal O}(e^{2\f})$ with
respect to $\Scl$, we are bound to obtain the action \uxvi\
in the end: there is only one conformally invariant action
(with a standard free field kinetic term for two fields
of opposite signature) that reduces in the weak coupling
limit to the classical dilaton-gravity action.

\subsection{The RST variant}

So far we started with a given kinetic part of the action
and improved the cosmological constant operator by higher
order corrections until it had dimension $(1,1)$. Russo,
Susskind and Thorlacius (RST)
\REF\RST{J. Russo, L. Susskind and L. Thorlacius,
\PR D46 1992 3444 .}
\REF\CC{J. Russo, L.
Susskind and L. Thorlacius,
 \PR D47 1993 533 .}
[\RST,\CC],
motivated by the search of simple and exactly solvable
\eoms, followed a slightly different route. Their procedure
amounts to keeping the cosmological constant operator fixed
and modifying the kinetic part of the action and hence also
the stress tensor until the old cosmological constant
operator has dimension $(1,1)$ with respect to the new
stress tensor. It turned out that this was very easy to
achieve. All one needs is to add
$$\delta S_{\rm RST}=-{\k\over 4\pi} \is \rg \f R = {1\over
\pi}\is\, \k\, \dpl\f\dm\r \ ,
\eqn\uxxii$$
where the second expression is valid in conformal gauge only.
In this RST-model the field transformations to the
$\x,\O$-fields are simplest:
$$\eqalign{\O&={e^{-2\f}\over \k}+{\f\over 2}\ ,\cr
\x&={e^{-2\f}\over \k}-{\f\over 2}+\r\ .}
\eqn\uxxiii$$
Again, when written in terms of these fields, the action
and stress tensor take on the form \uxvi\ and \uxii.
Note also, that for $\k>0$ the modified kinetic action is
degenerate at $e^{-2\f}=e^{-2\f_c}\equiv {\k\over 4}$. This
is precisely the value of $\f$ where  ${\rm d}\O/{\rm d}\f
=0$.  Although the precise value of $\f_c$ is shifted, the
qualitative feature of a singular line for $\k>0$ is
present in all models discussed so far.

\subsection{The de Alwis models}

More generally, one can add higher order corrections to
both the kinetic part of the action and to the cosmological
constant term. The final action $S[\x,\O]$ and stress
tensor $T_{\pm\pm}[\x,\O]$ are alway the same, but the
transformations between $\f,\r$ and $\x,\O$ are different.
This program was carried out by de Alwis
\REF\ALWI{S. de Alwis, \PR D46 1992 5429 .}. The maybe
somewhat unexpected result is that there are theories where
the transformations are one to one, i.e. have no singularity,
even for $\k>0$. We refer the reader to de Alwis' article
[\ALWI] for details.

\section{Singularity and shock-wave scenario}

As already repeatedly emphasized, in most models (e.g.
[\BC], [\RST]) for $\k>0$, the kinetic part of the
$\f,\r$-action is degenerate at some $\f=\f_c$. This
translates into a singularity of the transformation function
$\O(\f)$: ${{\rm d}\O\over {\rm d}\f}(\f_c)=0$. Since the
scalar curvature $R$ is proportional to $\left( {{\rm
d}\O\over {\rm d}\f} \right)^{-2}$, in general\foot{
In the RST-model, the only exception is the LDV
where $R=0$. }, the scalar curvature diverges on the line
where $\f=\f_c$. As shown first by RST [\RST, \CC], when
this line of singularity is time-like, one can impose
appropriate boundary conditions to avoid the curvature
singularity at $\f=\f_c$ and match the solution to
the LDV configuration.

A typical example is the shock-wave scenario for $\k>0$. Here
we only give a very short description. We refer the reader to
the RST-paper [\RST] for any details. One has the LDV for all
$\sp<\spo$. Then at $\sp=\spo$ a matter shock-wave,
characterised by $T_{++}=m\delta(\sp-\spo)$, passes by,
modifying the configuration for $\sp>\spo$. To
specify the initial data completely, we have also to give the
configuration on ${\cal I}_R^-$. We simply assume LDV
asymptotics. Then, the solution is such that a space-like
line of singular curvature extends from the shock-wave
trajectory into the region $\sp>\spo$, but it is hidden
behind an apparent horizon\foot{It is defined as the line
where $\dpl\f$ or equivalently $\dpl\O$ vanishes.} (which
starts as a time-like line). This is interpreted as the
formation of a black hole. Due to the emission of Hawking
radiation (signalled by a non-vanishing $T^M_{--}$), the
apparent horizon recedes, and intersects the line of
singularity in a finite proper time at $(\sps,\sms)$.
Beyond this point, the singularity turns time-like and
naked. One cannot evolve the field equations past this
singularity, and one has to impose boundary conditions on
the singular line. RST observed that on the future part of
the null-line going through $(\sps,\sms)$ the fields have
values precisely such that one can match them continuously
to a LDV configuration in the causal future of $(\sps,\sms)$
(at the cost of a delta-function type singularity in the
second derivatives, and hence in the stress tensor: this
gives rise to a ``thunderpop", an emission of energy of
order $\k\l$). This avoids the naked time-like curvature
singularity, and, maybe more important, it also stops the
Hawking radiation at $\sms$. This is physically important,
since at $\sm=\sms$ the initially formed black hole has lost
almost all of its initial mass $m$, and if Hawking radiation
would continue, one would inevitably arrive at
configurations of more and more negative total energies.
Thus the RST boundary condition serves to stabilize the
ground state of the system. Of course, the same applies to
the original model of ref. [\BC] discussed above.

\section{A local version of the covariant anomaly}

For various purposes we need to rewrite the covariant
anomaly $R{1\over \nabla^2}R$ in a local form. Consider
\REF\ST{L. Susskind and L. Thorlacius, \NP B382 1992 123 .}
[\ST]    %
$$S_Z={1\over 2\pi} \ix \rg\left[ -\half (\nabla Z)^2
+QRZ\right] \ .
\eqn\uxxiv$$
If we write
$$Z=\zt - Q{1\over \N^2}R\ ,\quad 2Q^2=\k\ ,
\eqn\uxxv$$
we have
$$S_Z=-{1\over 4\pi}\ix\rg (\N \zt)^2 - {Q^2\over 4\pi}
\ix\rg R{1\over \N^2} R\ ,
\eqn\uxxvi$$
which is a free-field action for $\zt$ plus $\Sanom$.
However, $\zt$ still ``remembers" the curvature coupling of
the original $Z$ field, since its stress tensor in
conformal gauge ($Z=\zt + 2Q\r$) reads
$$T_{\pm\pm}^\zt =\half (\d_\pm \zt)^2+Q\d_\pm^2\zt\ .
\eqn\uxxvii$$
and $T^Z_{\pm\pm}=T_{\pm\pm}^\zt+T_{\pm\pm}^{\r,{\rm
anom}}$, where $T_{\pm\pm}^{\r,{\rm anom}}= -\k \left(
(\dpl\r)^2-\dpl^2\r\right)$.
We see that $T_{\pm\pm}^\zt $ has a classical central
charge equal to $12\k=N-26+1+1$, hence it really represents
the contribution of the $N$ matter fields, the ghosts and
certain quantum fluctuations of $\f$ and $\r$. Thus, at the
semiclassical level, where we only consider the $\r$ and $\f$
(or $\x$ and $\O$) \eoms and the constraint equations
$$T_{\pm\pm}=T^{\f,\r,{\rm cl}}_{\pm\pm}+T^Z_{\pm\pm}=
T^{\f,\r}_{\pm\pm}+T_{\pm\pm}^\zt\ ,
\eqn\uxxviii$$
studying $\Scl+\Sanom+\delta S_{\rm improvement}$ is
completely equivalent to studying
$$S=S_{\rm cl}^{\rm no\ matter}+\delta S_{\rm improvement}
+S_Z\ .
\eqn\uxxix$$
Here $\delta S_{\rm improvement}$ stands for whatever
higher order corrections we added to the classical dilaton
gravity action. For the RST variant e.g. we have
$$S={1\over 2\pi}\ix \rg\left[ \left( e^{-2\f}-{\k\over
2}\f\right) R+e^{-2\f}\left( 4(\nabla\f)^2+4\l^2\right)
-\half (\nabla Z)^2 +QRZ\right]\ .
\eqn\uxxx$$
Note that by eq. \uxxv, the field $Z$ will be real only if
$\k>0$.

{\chapter{ The supersymmetric extension}}

\REF\PS{Y. Park and A.
Strominger, \PR D47 1993 1569 .}
A supersymmetric extension of the CGHS model was constructed
by Park and Strominger [\PS].
It seems natural to expect that the exactly solvable
quantum-improved theories discussed in the previous section
also have supersymmetric extensions. In fact there are
three different problems one might consider:
\pointbegin
Find a generally covariant supersymmetric extension of the
exactly solvable quantum-corrected actions, e.g. of the RST
action \uxxx. By supersymmetric extension one means a
supersymmetric action that in its bosonic sector
(setting all fermions equal to zero, and replacing the
auxiliary fields by the solutions of their algebraic field
equations) reduces  to the exact conformal, exactly solvable
quantum-corrected action.
\point
Find an exact superconformal theory that reduces in its
bosonic sector to the exact conformal, exactly solvable
quantum-corrected theory under consideration.
\point
Find an exact superconformal theory that reduces in its
bosonic sector and in the weak-coupling limit ($e^{2\f}\to
0$) to the classical dilaton-gravity theory (not
conformally invariant).
\par

Obviously, problem 3 is a weaker version of problem 2. It
might not be obvious at first sight why problem 1 and 2
should be different. However, problem 1 was solved in
\REF\BILPOS{A. Bilal, ``Positive energy theorem and
supersymmetry in exactly solvable quantum-corrected 2D
dilaton-gravity", Princeton University preprint PUPT-1373
(January 1993, revised April 1993), hep-th@xxx/9301021, to
appear in Phys. Rev. D48.}
ref. [\BILPOS]. On
the other hand, problem 2 has no solution as shown by
Nojiri and Oda
\REF\NOJ{S. Nojiri and I. Oda, \MPL A8 1993 53 .} [\NOJ].
Problem 3 was solved by Danielsson
\REF\DAN{U. Danielsson, \PL B307 1993 44 .}
[\DAN]
who also explained why problems 1 and 2 are different.
Here, we will discuss the solution to problem 1 and
then show why problem 2 cannot have a solution.

Start from a general supersymmetric dilaton-gravity
action in 2D
$$S^{(1)}={i\over 2\pi}\ixt\, E\left[
J(\Phi)S+iK(\Phi)D_\alpha \Phi D^\alpha \Phi + L(\Phi)
\right]
\eqn\qi$$
($\Phi$ is the dilaton superfield, $S$ the survature
multiplet and $E$ the super-zweibein, see ref. \PS\ for
all notation and conventions). $J, K$ and $L$ are for the
moment arbitrary scalar functions of the dilaton superfield.
If we expand the superfields, set all fermionic fields to
zero and integrate out the auxiliary fields (i.e. replace
them with the solutions of their algebraic field equations)
we get the  bosonic part of the action \qi\ [\PS]
$$S^{(1)}_{\rm bos}={1\over 2\pi}\ix\rg\left[ J\, R + 2K\,
(\N\f)^2 +\left({LL'\over 2J'}-{KL^2\over
2J'^2}\right)\right]\ ,
\eqn\qii$$
where now $J=J(\f), K=K(\f), L=L(\f)$ are functions of the
dilaton field. Choosing e.g. $J=e^{-2\f},\ K=2e^{-2\f},\
{LL'\over 2J'}-{KL^2\over 2J'^2}=4\l^2 e^{-2\f}$ with
solution $L=\pm 4\l e^{-2\f}$ reproduces $\Scl$.

We now repeat this exercise, including a supersymmetric
$Z$-field:
$$S^{(2)}={i\over 2\pi}\ixt E\, \left[
-{i\over 4}D_\alpha {\cal Z} D^\alpha {\cal Z} + Q{\cal
Z}S \right]\ .
\eqn\qiii$$
The bosonic part of this action alone is just $S_Z$ of
eq. \uxxiv.
When combining $S^{(1)}$ and $S^{(2)}$, the auxiliary
field equations get modified and the resulting bosonic
part is not just \qii\ plus $S_Z$, but rather
$$\eqalign{
\left[S^{(1)}+S^{(2)}\right]_{\rm bos}
&={1\over 2\pi}\ix\rg\left[ J\, R +
2K\,  (\N\f)^2 -\half (\N Z)^2+QRZ + F(\f)\right] \ ,\cr
F(\f)&=\left(1-{2\k K\over J'^2}\right)^{-1}
\left({LL'\over 2J'}-{KL^2\over
2J'^2}-{\k L'^2\over 4 J'^2}\right)\ .\cr }
\eqn\qiv$$
All we have to do now is to identify the functions $J, K$
and $L$ of $\f$ that reproduce e.g. the RST-action \uxxx.
For the latter we need
$$J(\f)=e^{-2\f}-{\k\over 2}\f\ ,\quad
K(\f)=2e^{-2\f}\ ,\quad
F(\f)=4\l^2 e^{-2\f}\ .
\eqn\qvi$$
Substituting these into the equation \qiv\ for $F$ we obtain
a non-linear differential equation for $L(\f)$ :
$$(L+xL')(L+L')=-\k^2\l^2 {(1-x)^2\over x^2}
\eqn\qvii$$
where $L'={\rm d}L/{\rm d}\f$ and $x={\k\over 4}e^{2\f}$.
The solution is very simple:
$$L(\f)=\pm4\l\left( e^{-2\f} +{\k\over 4}\right) \ .
\eqn\qviii$$
Obviously there are two choices of sign since only $\l^2$
is relevant. Thus, if $J, K$ and $L$ are given by \qvi,
\qviii, the action $S^{(1)}+S^{(2)}$ is a supersymmetric
extension of the RST-action.

Similarly, we can construct a supersymmetric extension of
the action of ref. \BC. In this case $J(\f)$ and $K(\f)$
are given by the CGHS-functions
$$J(\f)=e^{-2\f}\ , \quad K(\f)=2 e^{-2\f}
\eqn\qix$$
while the function $F$ is more complicated [\BC]. As a
consequence, the non-linear differential equation to be
solved for $L(\f)$ is considerably more involved. After some
exercise  (see  ref. [\BILPOS] for details), one finds
$L(\f)$ as a transcendental function of $\O(\f)$. Here we
only give its weak-coupling expansion
 for small $\k e^{2\f}$
$$L(\f)\sim \pm 4\l \left( e^{-2\f} +{\k\over 4} \f +\tilde
c\right)\ .
\eqn\qxvi$$
Note that, as expected, the leading term in the
weak-coupling expansion of $L(\f)$ for both variants
discussed here is $\pm 4\l  e^{-2\f}$ which is the $L(\f)$
as appropriate for $\Scl$. We also note that $L$ is linear
in $\l$ and that the cosmological term $F$ in the bosonic
part obtained after solving the auxiliary field equations is
bilinear in $L$, hence $\sim \l^2$ as it should.

Now what is the problem with problem 2 ? As just pointed
out, upon integrating out the auxiliary fields, one
replaces operators linear in $L$ by operators bilinear in
$L$. These bilinears have to be regularized, so that the
naive procedure of integrating out the auxiliary fields can
only be trusted at the semiclassical level [\DAN].

In other
words: at the (semi)classical level, if $L$ has
classical conformal dimension $({1\over 2},{1\over 2})$
then $L^2$ has dimension $(1,1)$ and vice versa. However,
quantum mechanically, due to the necessity to
regularize $L^2$, these two statements are in general
incompatible. The requirement that the bosonic part be an
exact conformal theory means that $L^2$ has to have
dimension $(1,1)$ quantum mechanically. Exact superconformal
invariance means that $L$ has to have dimension $({1\over
2},{1\over 2})$ quantum mechanically. We can arrange for one
or the other but not for both at the same time. Problem 2
consists of achieving both which is impossible (or rather has
only the trivial solution $L=0$ [\NOJ]). Problem 1 insists on
$L^2$ having dimension $(1,1)$, so the susy theory is not an
exact superconformal theory. Finally, problem 3 insists on
$L$ having dimension $({1\over 2},{1\over 2})$, and thus the
bosonic part is not an exact conformal theory, but can still
be arranged to have the correct weak-coupling classical
limit [\DAN].

\chapter{Canonical formalism, ADM energy and functional
integral measure}

We would like to display a canonical formalism for the
dilaton-gravity theories in 2D. The starting point is the
covariant, i.e. not gauge-fixed theory, and hence we
represent the anomaly term by the local and covariant
$Z$-action. We can treat the classical model, the CGHS
model and the RST model simultaneously if we consider the
following action (cf. \uxxx)
$$
S = {1\over 2\pi} \is \rg
\left[ \left( e^{-2\f} -{\k\over 2}\f\right) R
+4 e^{-2\f} \left( (\nabla\f)^2 + \l^2\right)
-\half (\nabla Z)^2 +QRZ\right]
\eqn\vii$$
where $\k=Q=0$ gives back the classical action (with one
free matter $Z$-field\foot{
One might add other free (classical) matter fields. These
could be included trivially into  our
subsequent analysis, in particular their contribution to
the boundary term $D$ would vanish due to the standard
boundary conditions on matter fields.}), $\k=0,\ 2Q^2={N\over
12}$ gives the CGHS model [\CGHS], and $\k=2Q^2={N-24\over
12}$ gives the RST-model [\RST].

\def\d{\delta}
\def\pa{\partial}
\def\ds{{\rm d} \sigma}

\section{Canonical formalism}

We parametrize the
two-dimensional metric in the following way
$$g_{\m\n}=e^{2\r}\ \pmatrix{ A^2-B^2 & A\cr A & 1\cr}\ .
\eqn\di$$
This is inspired by the standard ADM parametrization
\REF\ADM{R.L. Arnowitt, S. Deser and C.W. Misner, in
``Gravitation: An introduction to current research", ed. L.
Witten, Wiley, New York, 1962;
see also: C.W. Misner, K.S. Thorne and J.A. Wheeler,
``Gravitation", W.H. Freeman \& Company, San Francisco,
1973. }
[\ADM] with $A$ and $e^\r B$ the analogues of the
shift vector and lapse function. Conformal gauge is simply
$A=0,\ B=1$. Inserting this parametrization into the above
action we obtain $S[A,B,\f,\r,Z,\dot\f,\dot\r,\dot Z]$
which does not depend on $\dot A$ or $\dot B$. We refer the
reader to
\REF\BK{A. Bilal and I. Kogan, \PR D47 1993 5408 .}
ref. [\BK] for details.  It is straightforward to compute
the momenta $\pf, \pr$ and $\pz$. Their general expression
can be found in [\BK]. Here we only give them for $A=0,
B=1$ (conformal gauge)
$$
\pf={1\over \pi }\left( F \dot\r
-2e^{-2\f}\dot\f\right)\ ,\quad
\pr={1\over \pi }\left( F \dot\f
-{Q\over 2}\dot Z\right)\ ,\quad
\pz={1\over 4\pi }\left( \dot Z
-2Q \dot\r\right)\ ,
\eqn\diii$$
where $F=e^{-2\f}+{\k\over 4}$.
Since no time-derivatives of $A$ or $B$ occur in the action
there are no momenta conjugate to $A$ or $B$. The fields
$A$ and $B$ are Lagrange multipliers serving to impose
constraints.

Writing $S=\int {\rm d}\t L$, the bulk
Hamiltonian is given by $H_1=\int \ds (\dot\f\pf+\dot\r\pr+
\dot Z\pz )-L$ which after integrating by parts reads
$$\eqalign{
H_1=&\int \ds [ A\ca +B\cb ]\cr
\ca=&\f'\pf+\r'\pr+Z'\pz -\pr'\cr
\cb=&{\pi\over G^2}\left[
e^{-2\f}(\pr+2Q\pz)^2+F\pf(\pr+2Q\pz)+{1\over
2}Q^2\pf^2\right] +2\pi\pz^2\cr
&+{1\over \pi}\left[ F\r'\f'-(F\f')'-e^{-2\f}\f'^2-
\l^2e^{-2\f+2\r}+{1\over 8}Z'^2-{1\over 2}Q\r'Z'+
{1\over 2}QZ''\right]}
\eqn\div$$
where $G^2=F^2-2Q^2e^{-2\f}$.

The Lagrange multipliers $A$ and $B$ impose the constraints
$\ca=\cb=0$. What is the interpretation of these
constraints? We evaluate them in conformal gauge
($A=0,B=1$) and substitute \diii\ for the momenta. Then
(using also $Z=\zt+2Q\r$)  $\ca$ and $\cb$ are easily seen
to coincide with  ${1\over 2\pi}(T_{++}-T_{--})$ and
${1\over 2\pi}(T_{++}+T_{--})$, respectively.

Using canonical Poisson brackets,
$$\{\f(\s),\pf(\s')\}=
\{\r(\s),\pr(\s')\}=
\{Z(\s),\pz(\s')\}=\d (\s-\s')
\eqn\dvi$$
we can compute the algebra of the constraints as given by
\div\ (in general gauge). We find
 that the Poisson bracket of $\ca+\cb$ with
$\ca-\cb$ vanishes while
$$
\{ (\ca(\s)\pm \cb(\s)),(\ca(\s')\pm \cb(\s')) \}
=2(\pa_\s-\pa_{\s'})[(\ca(\s')\pm \cb(\s'))\d(\s-\s')]
\eqn\dviii$$
which is indeed the Poisson bracket algebra of $T_{\pm\pm}$
with itself.  There is no $\d'''$-term which means that
the total central charge vanishes. This was to be expected:
for the classical theory this is obvious, while for the
RST-model e.g. $Q^2(\pa_\pm\r\pa_\pm\r -\pa_\pm^2\r)$ gives
$c=-24Q^2=-12\k=24-N$, and the $\zt$-field gives the
anomaly for matter, ghosts and the quantum part of $\f,\r$
which is $c=N-26+2=N-24$. Of course, we just repeated that
the Polyakov-anomaly action is designed to cancel the
various anomalies present in the theory.

According to the variational principle in Hamiltonian form,
$$\delta \left( \int\ds
(\dot\f\pf+\dot\r\pr+\dot Z\pz)
- H_1\right)=0$$
 should be
equivalent to Hamilton's equations, ${\d H_1\over
\d\varphi_i}+\dot\Pi_i={\d H_1\over
\d\Pi_i}-\dot\varphi_i=0$.
However, if we carefully carry out the variation, keeping
track of boundary terms arising from integrating by parts,
we find
$$\delta \left( \int\ds
(\dot\f\pf+\dot\r\pr+\dot Z\pz) - H_1\right)=-D+X\ ,
\eqn\dx$$
where $D$ is a boundary term and the vanishing of $X$
is precisely equivalent to Hamilton's equations.
For $A=0,B=1$ (conformal gauge), the boundary term $D$ is
given by (recall that $F=e^{-2\f}+{\k\over 4}$)
$$\eqalign{ \pi D = \Bigg[
&\d\left(-F\f'+{1\over 2}QZ'\right) +
\left( F\r'-2e^{-2\f}\f'\right)\d\f\cr
&+\left( F\f'-{1\over 2}QZ'\right)\d\r
+\left( -{1\over 2}Q\r'+{1\over 4}Z'
\right)\d Z \Bigg]^{\sep}_{\sem}\ .\cr}
\eqn\dxii$$
This does not vanish.

\section{The ADM energy}

The same situation occurs in classical general relativity
in four dimensions. The ADM Hamiltonian  gives the \eoms
only up to boundary terms. As first noticed by
Regge and Teitelboim
\REF\RTE{T. Regge and C. Teitelboim, Ann. Phys. {\bf 88}
(1974) 286.} [\RTE]
for asymptotically flat space-times this has a very
simple resolution. The idea is to take these boundary terms
serious and show that under appropriate asymptotic
conditions on the fields, the boundary term can be canceled
by the variation of an appropriate  boundary Hamiltonian
$H_2$ added to the bulk Hamiltonian $H_1$. The sum
$H=H_1+H_2$ then is interpreted as the true Hamiltonian.
Since $H_1$ is only a sum of constraints, its numerical
value on any solution vanishes, and the total energy, i.e.
the numerical value of $H$ is given by that of $H_2$. Thus
the total energy is automatically given by the boundary
terms.

In the present 2D dilaton-gravities the same trick works.
Here, we only discuss the case $A=0,B=1$ and refer
to [\BK] for the general case. As in 4D, we assume
asymptotically flat space, and asymptotically Minkowskian
coordinates. This is translated into requiring linear
dilaton vacuum asymptotics:
$${\rm as\ } \s\to\pm\infty\ :\quad \f\sim -\l\s\ ,\ \r\sim
0\ ,\ Z\sim 0\ ,
\eqn\dxiii$$
(and corresponding conditions on the momenta although in
conformal gauge they are not really needed). For $\s\to
+\infty$ we have to add another coordinate condition:
$$
{\rm as}\ \stp\ :\quad e^{2\l\s} (\f+\l\s-\r)\to 0\ .
\eqn\dxiv$$
One can indeed convince oneself that this can be satisfied
for all models that differ from $\Scl$ only by terms ${\cal
O}(e^{2\f})$. We now restrict the phase-space to only those
configurations that obey the asymptotic conditions \dxiii\
and \dxiv. This is a perfectly legitimate procedure.
Equation \dxiv\ implies in particular that $\lim_{\stp}
e^{-2\f}\d\f = \lim_{\stp} e^{-2\f}\d\r$. Using these kinds
of relations, one easily sees that
$$ D\Big\vert_{\rm boundary\ conditions} =-\d H_2\ ,\quad
H_2= {1\over 2\pi} \left[ e^{-2\f}\left( 2 \f' +
\l\right) +\l e^{2\l\s} \right] \Bigg\vert_{\sep}\ ,
\eqn\dxv$$
where we adjusted an (infinite) additive field-independent
term, not affecting the relation $\d H_2=-D$, so that $H_2$
vanishes for the LDV. Note that as a consequence of the
boundary conditions, $H_2$ receives no contribution from
$\sem$.
Then
$$\d \left( \int\ds
(\dot\f\pf+\dot\r\pr+\dot Z\pz) - H_1-H_2\right)=0
\quad \Rightarrow \quad {\rm Hamilton's\ equations}\ .
\eqn\dxvi$$
Thus the true Hamiltonian is $H=H_1+H_2$. Since
$H_1$ vanishes on all solutions, the total energy is given
by the value of $H_2$ only (times a conventional
normalization factor of $2\pi$).
If we
write $\f\sim -\l\s+\d\f$ with $\d\f$ at least ${\cal
O}(e^{-\l\s})$ then we can write the total energy $M=2\pi
H$ as
$$M=  \lim_{\stp} 2e^{2\l\s} (\pa_\s +\l) (\d\f -\d\f^2)
\eqn\dxvii$$
Due to our boundary conditions we can rewrite the part linear
in $\d\f$ as
$$M^{(1)}=  \lim_{\stp} 2e^{2\l\s} (\pa_\s\d\f +\l\d\r)
\eqn\dxviii$$
which is the formula for the ADM-mass usually given in the
literature
\REF\WIT{E. Witten, \PR D44 1991 314 .}
[\WIT]. However this latter formula is only correct if the
$\d\f^2$ term can be neglected, i.e. if $\d\f={\cal
O}(e^{-2\l\s})$ as $\stp$. Indeed, in general the $\d\f^2$
term is crucial to make $M$ time independent. Using \dxiv\ it
is easy to see that the full expression $M$ as given by
\dxvii\ obeys
$$ {{\rm d}\over {\rm d}\t} M= -2 \pi \ca\Bigg\vert_{\sep} \
, \eqn\dxix$$
and since $\ca=0$ is a constraint,
${{\rm d}\over {\rm d}\t} M$ vanishes.

Note that the same mass formula applies to
all three models, classical, CGHS and RST.

\section{A note on the functional integral measure}
\def\D{{\cal D}}
\def\po{\Pi_\O}
\def\px{\Pi_\x}

In a functional integral approach, any transformation on
the fields is accompanied by a Jacobian for the measure.
One might wonder whether we should worry about such a
Jacobian when we transform from $\f,\r$ to $\O,\x$. I shall
now argue that the final functional integral where one
integrates over $\O$ and $\x$ contains no Jacobian
\REF\BKD{These remarks are based on discussions with I.
Kogan.} [\BKD]. The important point is that it is {\it
incorrect} to start with a functional integral ${\cal
Z}=\int \D\f\D\r \exp\left\{-S[\f,\r]\right\}$. Of course, we
know that the correct starting point for the functional
integral is
$${\cal Z}=\int \D\f\D\r\D\pf\D\pr
\exp\left\{-\int
\left(\dot\f\pf+\dot\r\pr\right)+\int
H[\f,\r,\pf,\pr]\right\}\ .
\eqn\dxx$$
It is only when the Hamiltonian $H$ has a standard kinetic
term, $H=\sum_i a_i \Pi_i^2 +\ldots$ with constant $a_i$,
that one can perform the gaussian integration over the
momenta and obtain a functional integral over the fields
only. For the dilaton-gravity theories we are considering
this is certainly not the case, see e.g. eq. \div. The
Hamiltonian is still quadratic in the momenta, but the
coefficients are complicated $\f$-dependent functions.

Instead of trying to evaluate the functional integral \dxx\
we make the {\it canonical} transformation
$$\eqalign{
\f,\r\ &\to \ \O,\x\cr
\pf,\pr\ &\to \ \po,\px \ .}
\eqn\dxxi$$
The only things that matter is that the kinetic part of the
action is a free field action in terms of $\O$ and $\x$,
and that the transformation is canonical, which determines
$\po$ and $\px$. For the RST variant we have e.g.
$$\eqalign{
\po&=-{ \left( e^{-2\f}+{\k\over 4}\right) \pr
+{\k\over 2}\pf \over
e^{-2\f}-{\k\over 4}} = {\k\over 2\pi} \dot\O\ ,\cr
\px&=\pr=-{\k\over 2\pi}\dot\x \ .\cr }
\eqn\dxxiaa$$
Now, on the one hand, since the transformation is canonical
$$\eqalign{
&\int \left(\dot\f\pf+\dot\r\pr\right)-H[\f,\r,\pf,\pr]\cr
=&\int \left(\dot\O\po+\dot\x\px\right)-H[\O,\x,\po,\px]\cr
=&\int \left(\dot\O\po+\dot\x\px\right)
-\int \left[ {\pi\over \k}\left( \po^2- \px^2 \right)
+{\k\over 4\pi}\left( (\O')^2 -  (\x')^2 +2\x'' \right)
-\l^2 e^{2(\x-\O)}  \right] \ . }
\eqn\dxxii$$
Indeed we see that variation of the latter yields the
$\O,\x$ \eoms in Hamiltonian form (up to boundary terms, see
below). On the other hand, a canonical transformation
preserves the phase space measure ${\rm d}q{\rm d}p$. Since
our transformation is local and contains no space-time
derivatives it is a canonical transformation at any point
$(\t,\s)$. Hence it certainly preserves the discretized
measure for the functional integral:
$$\eqalign{
&\prod_i \ {\rm d}\f(\t_i,\s_i){\rm d}\r(\t_i,\s_i)
{\rm d}\pf(\t_i,\s_i) {\rm d}\pr(\t_i,\s_i) \cr
&=\prod_i \ {\rm d}\O(\t_i,\s_i){\rm d}\x(\t_i,\s_i)
{\rm d}\po(\t_i,\s_i) {\rm d}\px(\t_i,\s_i)\ .\cr}
\eqn\dxxiii$$
Thus we expect this equality
to hold in the continuum limit,
$$\D\f\D\r\D\pf\D\pr = \D\O\D\x\D\po\D\px\ ,
\eqn\dxxiiia$$
and hence
$${\cal Z}=\int \D\O\D\x\D\po\D\px
\exp\left\{- \int \left(\dot\O\po+\dot\x\px\right)
+\int H[\O,\x,\po,\px] \right\}\ ,
\eqn\dxxiv$$
where the argument of the exponential is given by the
negative of \dxxii. It is now of the standard quadratic form
$\dot q p- a p^2$ with constant $a$, and we can perform the
gaussian integration with the result
$${\cal Z}=\int \D\O\D\x
\exp\left\{- S[\O,\x] \right\}\ ,
\eqn\dxxv$$
where $S[\O,\x]={1\over \pi}\is  \left[ -\k\dpl\x\dm\x+\k
\dpl\O\dm\O + \l^2 e^{2(\x-\O)}  \right] $,
cf. eq. \uxvi\ (up to a boundary term). We
see that this functional integral contains no extra
Jacobian. If we however change variables back to $\f,\r$, we
get a Jacobian, so that the correct $\f,\r$ functional
intagral is ${\cal Z}=\int \D\f\D\r\ J\ e^{- S[\f,\r]}$,
where $J={\pa(\O,\x)\over \pa(\f,\r)}$.

As it stands, the functional integral \dxxv\ does not take
into account boundary effects. In particular, we have seen
in the preceeding subsection that one has to add a boundary
Hamiltonian $H_2$ so that the solutions to the \eoms
provide a true saddle point to the functional integral. For
the RST variant e.g., we have from eq. \dxxii\ $D={\k\over
2\pi} \left[ \O'
\d\O-\x'\d\x+\d\x'\right]\Big\vert^{\sep}_{\sem}$, and
using the same boundary conditions as above one has $D=-\d
H_2$ with the boundary Hamiltonian $H_2=\left[ {\k\over 2\pi}
(\l-\pa_\s )\O +{\l\over 2\pi} e^{2\l\s}\right]_{\sep}$. (Of
course, this coincides with eq. \dxv.)
It should not be
too difficult to impose the appropriate boundary conditions
on the fields in the functional integral. Thus, for $\k<0$,
one should be able to quantize the theory using the
functional integral. For $\k>0$ however, we have the extra
restriction $\O>\O_c$ on  the range of integration of $\O$.
At present, nobody seems to know how to evaluate such
functional integrals.

\chapter{A positive energy theorem}

\section{The theorem}

We will give a positive energy theorem for the RST variant
of the exactly solvable dilaton-gravity theories. Although
its formulation involves some spinors, it only relies on
the (bosonic) \eoms, and the spinors are simply a
convenient device to express certain dependences on the
dilaton and metric fields. Of  course, the spinorial
formulation is  motivated by the existence of a supercharge
in the supersymmetric extension, and the deep reason why we
can prove a positive energy theorem is probably the
existence of this supersymmetric extension. Nevertheless, let
us stress again that in this section we are dealing with a
purely bosonic theory. We will need the
\eoms for the metric in the covariant form, $T_{\m\n}=0$.
Thus we will again make use of the reformulation \uxxx\
using the $Z$-field. The proof we will give [\BILPOS]  for
the RST model
 is a generalization of the one given by Park and
Strominger [\PS] for the CGHS model. We will assume $\k>0$
throughout this section, so that $Z$ is real.

The \eoms of the metric are
$T_{\m\n}^{g,\f}+T_{\m\n}^Z=0$,
where $T_{\m\n}^Z$ is obtained from the action \uxxiv:
$$\eqalign{\tmn^Z&=\hat \tmn^Z +Q\left( \N_\m\N_\n
Z-g_{\m\n}\N^2Z\right)\ ,\cr
\hat \tmn^Z &={1\over 2} \N_\m Z \N_\n Z-{1\over 4}
g_{\m\n}(\N Z)^2\ ,}
\eqn\tia$$
and
$T_{\m\n}^{g,\f}$ is the covariant form of
$T_{\pm\pm}^{\r,\f}-T_{\pm\pm}^{\r,{\rm anom}}$, namely
$$\eqalign{T_{\m\n}^{g,\f}=&-2\left( e^{-2\f}+{\k\over
4}\right) \left( \N_\m\N_\n\f-g_{\m\n}\N^2\f\right)\cr
&-2 e^{-2\f}g_{\m\n}\left( (\N\f)^2-\l^2\right)\ .}
\eqn\tii$$
Note that we have decomposed $T_{\m\n}^Z$ into a piece
$\hat \tmn^Z$ that does obey the dominant energy condition
(it is just the standard stress energy tensor for a free
matter field) and a piece $\sim Q$ that does not. We then
have the following

{\it THEOREM}: Let
$$M=\int_\S {\rm d}\s^\m \, \N_\m\left[
2\left( e^{-2\f}+{\k\over 4}\right)
\eb\gf(\NS\f-\l)\e-Q\eb\gf\NS Z\e\right]\ ,
\eqn\tiii$$
where $2Q^2=\k>0$. $\e$ is a commuting real two-dimensional
spinor, and $\NS=\g^\m\N_\m=e^{\m a}\G_a\N_\m$, $e^{\m
a}$ being the zwei-bein. \foot{The $\G_a$ are
Minkowski-space Dirac-matrices obeying
$\{\G_a,\G_b\}=2\eta_{ab}$. A convenient choice that we
adopt here is $\G_0=i\s_y,\, \G_1=\s_x$. Let
$\G_5=\G_0\G_1=\s_z$ while (following ref. \PS)
$\gf=\g^0\g^1=-\G_5$. As usual, $\eb=\e^+\G_0$. The
antisymmetric tensor is
$\e_{0\phantom{1}}^{\phantom{0}1}
=\e_{1\phantom{0}}^{\phantom{1}0}=-1$.}
Then
$$M\ge 0
\eqn\tiiia$$
if (i) the line $\S$ is space-like or null, (ii) $\f$ is
real on all of $\S$ (i.e. $\f\le\f_c$ or $\f$ is the LDV),
and (iii) $\e$ is a solution of the ordinary differential
equation on $\S$
$${\rm d}\s^\m \left[\left( 1+{\k\over 4}e^{2\f}\right)
\N_\m \e -\half \g_\m (\NS\f-\l)\e -{Q\over 4}
e^{2\f}\left( 1+{\k\over 4}e^{2\f}\right)^{-1} \g_\m \NS Z
\e\, \right] =0\ .
\eqn\tv$$
Note that the $\e$-differential equation determines $\e$
only up to two functions of integration. Thus $M$ not only
depends on the fields $\f,g_{\m\n}$ and the line $\S$, but
also on these functions of integration. We will fix the
latter in the next subsection.

{\it Proof} (sketched only, for more details, see
[\BILPOS]): One evaluates $\N_\m\left[ \ldots \right]$ in
\tiii. Whenever one has $\N_\m \e$ or $\N_\m \eb$ one uses
the $\e$-differential equation to eliminate it. Using some
trivial spinor identities, one can identify various pieces
of the stress energy tensor. Using the \eoms
$T_{\m\n}^{g,\f}+T_{\m\n}^Z=0$ it can be seen that $M$ equals
$$M=\int_\S {\rm d}\s^\m \left( {1-{\k\over 4}e^{2\f}\over
1+{\k\over 4}e^{2\f}}\right)^2
\e_{\m\phantom{1}}^{\phantom{0}\r}\, \hat T^Z_{\r\n} \,
\eb\g^\n\e \ .
\eqn\tiv$$
This expression  is manifestly
non-negative for $\k>0$ if $\f$ is real everywhere on $\S$.
Indeed, it is easy to see that for any real non-zero
$\e=\pmatrix{\e_1\cr\e_2\cr} $ (not necessarily a solution
of \tv), $v^\n=\eb\g^\n\e$ is time-like or null and
future-directed.  Now $\hat T^Z_{\m\n}$
obeys the dominant energy condition, i.e. for time-like or
null, future-directed $v^\n$ the vector $-\hat
T^Z_{\m\n}v^\n$ is again time-like or null, future-directed.
Note that this is true only if $Z$ is real, i.e. for $\k>0$!
 Since
$\e_{1\phantom{1}}^{\phantom{0}0}=-1$ it follows that $M$ as
given by \tiv\ is non-negative provided $\S$ is space-like
or null and $\f$ real on $\S$.

In ref. [\BILPOS] it was also shown that the functional $M$
as given by \tiii\ is unique in the following sense: Suppose
one replaces the coefficients of $\NS\f$, $\l$, $\NS Z$
(which are the only covariant objects of the dimension of a
mass) in \tiii\ and in \tv\ by arbitrary scalar functions of
$\f$. Then we will be able to use the \eoms to obtain a
non-negative quantity of the type \tiv\ involving the $\hat
T^Z$ (which obeys the dominat energy condition)  only if
the functions of $\f$ are precisely as given in \tiii\ and
\tv.

Although  the mass
{\it functional}  \tiii\ is uniquely determined, its actual
value depends on the boundary or initial conditions imposed
on the spinor $\e$ upon solving its differential equation
\tv. They will be fixed  next  by imposing
physical requirements.

\section{Physical interpretation and applications}

The functional $M$ is given by a line integral of a
derivative along this line and thus reduces to the
difference of the values of the expression in the square
brackets at ``both ends of the world". Thus $M$ is given by
the asymptotic values of the fields and of the spinor
$\e$. However, through the $\e$-differential equation the
latter depend on the fields on all of $\S$. This differs from
4D general relativity.

Now we would like to see
whether the non-negative functional $M$ defines a reasonable
mass (energy) and compute it for various physically
interesting scenarios.  In particular, we will evaluate $M$
as defined in \tiii\ for the case where the field
configuration is asymptotic to the LDV at both ends of $\S$.
If $\S$ is a space-like line one should obtain the ADM-mass
while a null-line $\S$ should lead to the Bondi-mass.
This has been discussed  in considerable detail in
[\BILPOS]. Here we will just give the main results. All
computations in this subsection will be in conformal gauge.

\subsection{$\S$ a space-like line of constant
$\t$: ADM-mass}

If we denote the
expression in square brackets in \tiii\ by $\M$ we  have
$$M(\t)=\M(\t,\s=\infty) - \M(\t,\s=-\infty)\ .
\eqn\txi$$
We assume LDV asymptotics,
i.e.  as $\s\to \pm\infty$ :
$\f\sim -\l\s+\delta\f\ , \ \r\sim \delta\r$
where $ \delta\f$ and $\delta\r$ vanish as $\s\to
\pm\infty$.  Let furthermore $\zt\to 0$
as $\s\to \pm\infty$ so that $Z\sim 2Q\r$.
More precisely, the LDV asymptotics as $\s\to +\infty$
together with the \eoms and constraints imply
$$\eqalign{
&\r\sim a_1(\t)e^{-\l\s}+a_2(\t)e^{-2\l\s}+\ldots\ ,\ \
\f=-\l\s+\r\cr
&\dot a_2=2a_1 \dot a_1\ \ , \quad \ddot a_1 =\l^2 a_1\
,\cr}
\eqn\qqiiia$$
where we use a suitable set of coordinates so that
$\f=-\l\s+\r$. It is related to the ``Kruskal" coordinates
$x^\pm$ where $\f=\r$ by the usual transformation
$\l x^\pm=e^{\pm\l\s^\pm}$.
Then the asymptotics of the differential equation for $\e$
implies%
$$
\e={c_0\over \sqrt{2}}\pmatrix{1\cr -1\cr}+
{c_1\over \sqrt{2}}\pmatrix{1\cr 1\cr}e^{-\l\s}
+ \Or (e^{-2\l\s})
\eqn\qqiv$$
where $c_0$ and $c_1$ may depend on $\t$ and are the two
functions of integration. Note that this implies
$(\NS\f-\l)\e\vert_{\sep}=0$ which ensures that
$\M(\t,\s=+\infty)$ and hence $M$ do not diverge for
configurations asymptotic to the LDV. Given $c_0$ and $c_1$,
the differential equation completely determines $\e$, and in
particular its asymptotics as $\s\to -\infty$.

For the LDV, it is easy to solve the
$\e$-differential equation exactly and show that
$$M_{\rm LDV}=0
\eqn\qqix$$
independent of the choice of  the
functions of integration $c_0$ and $c_1$. It is worthwile
noting however, that unless $c_1=0$ the total energy for
the LDV receives contributions from both ends of $\S$ (which
cancel each other). One sees that one has to take carefully
into account $\M(\t,\s=-\infty)$ as well as the subleading
term ($\sim \Or (e^{-\l\s})$) in $\e$ when evaluating
$\M(\t,\s=+\infty)$.

For the general LDV-asymptotic configuration \qqiiia\ we
have to fix the functions of integration $c_0$ and $c_1$.
We first impose
$$\
\eb\gf\e\vert_{\s=\infty} =1\ .
\eqn\qqxiib$$
which fixes
$c_0=1$. The other function of integration $c_1$ is
determined by requiring
$$
\lim_{\s\to\infty} e^{-\f} (\NS\f-\l)\e=0 \ .
\eqn\qqxiia$$
Indeed, this fixes the subleading term  from
the expansion \qqiv\ of $\e$, since the leading term
vanishes automatically by the differential equation.
Equation \qqxiia\ determines $c_1$ as $c_1 = {\dot a_1\over
2\l} c_0= {\dot a_1\over 2\l}$.
Remark, that an alternative choice would be to replace
\qqxiia\ by the following condition at $\s=-\infty$:
$(\NS\f-\l)\e\vert_{\s=-\infty}=0$. Then
$\M(\t,\s=-\infty)=0$ and $M=\M(\t,\s=+\infty)$. On the
other hand, $c_1$ then has to be obtained by solving the
$\e$-differential equation for all $\s$. This type of
approach will be used when we compute the Bondi-mass, but it
could also be carried out for the present discussion of the
ADM-mass.

With our choice \qqxiib\ and \qqxiia\ we obtain
$$\eqalign{
\M(\t,\s=+\infty)&=2\l (a_1^2- a_2) +
{1\over \l} (\dot a_1^2-\l^2 a_1^2) \ ,\cr
-\M(\t,\s=-\infty) &={\k\over 2} \l \eb\gf(1+\G_1)\e
\vert_{\s=-\infty} \ge 0\ .\cr}
\eqn\qqxiva$$
We note that, by equation \qqiv\ the terms $\Or (e^{-2\l\s})$
in the asymptotic expansion of $\e$ do not contribute
to $\M(\t,\s=+\infty)$.
Using \qqiiia\ we get
$${{\rm d}\over {\rm d}\t} \M(\t,\s=+\infty)=0\ .
\eqn\qqxivb$$
Thus we find that at least the contribution from
$\s=+\infty$ does not depend on time.

Let us compare
\qqxiva\ with the expression for the ``true"
total energy  we
 derived in the previous section: $M_{\rm true}
= \lim_{\s\to\infty} 2e^{2\l\s}
(\d_\s+\l)(\delta\f-\delta\f^2)$. From \qqiiia\ we have
$M_{\rm true}=2\l (a_1^2- a_2)$. This differs from
$\M(\t,\s=+\infty)$ only by $\Delta={1\over \l} (\dot
a_1^2-\l^2 a_1^2) $. Now by eq. \qqiiia,
$a_1=a_+e^{\l\t}+a_-e^{-\l\t}$, so that $\Delta=-4\l a_+a_-$
is a constant. In many situation, due to the initial
conditions, either $a_+$ or $a_-$  vanishes and so does
$\Delta$. If this is the case, $\M(\t,\s=+\infty)=M_{\rm
true}$. More generally we have proven that
$$0\le M =
M_{\rm true} +\Delta +{\k\l\over 2}
\eb(-\infty)\gf(1+\G_1)\e(-\infty) \ .
\eqn\txv$$
If $\Delta\ne 0$ this is not of much use to prove
positivity properties of $M_{\rm true}$, but if one can
show on general grounds that $\Delta$ has to vanish, one has
$M_{\rm true}\ge M_{\rm min}$ where $M_{\rm min}
=-{\k\l\over 2}
\eb(-\infty)\gf(1+\G_1)\e(-\infty) $ is some ``slightly"
(i.e. naively of order $\k\l$) negative amount of energy.
This suggests the interpretation that the true total energy
is non-negative except for some amount $\Or(\k\l)$, due to
quantum effects, which is bounded  by $M_{\rm min}$.

\subsection{$\S$ a null line of constant $\sm=\t-\s$:
Bondi-mass}

On can solve the $\e$-differential equation exactly in
terms of integrals of functions of $\f$ and $\r$. We
have in mind to study the shock-wave scenario or any other
scenario with $T^{\rm M}_{++}\ne 0$ over a finite
interval in $\sp$ only, where we have the exact LDV for
all $\sp<\spo$ with some finite $\spo$. Hence it is
convenient to fix the initial conditions for $\e$ in the LDV
region or on its boundary at $\sp=\spo$ (where $\r=0$):
$$
\e(\spo,\sm)\ =\ \pmatrix{ d_1(\sm)\cr d_2(\sm) \cr }
\eqn\qqxvi$$
i.e. $d_1(\sm)$ and $d_2(\sm)$ are our functions of
integration. Then the asymptotics as $\sp\to
-\infty$ are relatively easy to obtain, while those for
$\sp\to\infty$ are more involved. We will not give any
details here but refer the interested reader to [\BILPOS].
The outcome of the computation is that
the Bondi-mass equals
$$\eqalign{
M_{\rm B}(\sm)&= \M(\sm,\sp=+\infty)-\M(\sm,\sp=-\infty)\cr
&={2d_2^2(\sm)\over 1-{p\over \l}e^{\l\sm}}
\left[ m +\kk \l
\log\left( 1-{p\over \l}e^{\l\sm}\right)
+\kk p e^{\l\sm}\right] \cr
&-\l e^{-\l\sm} L^2(\sm)
+2(d_1+d_2)^2 \l \left( e^{\l(\spo-\sm)}+\kk \right)\ , }
\eqn\qqxxviii$$
where $m$ and $p$ are the total energy and momentum carried
by the infalling matter. (The shock-wave corresponds to
$p=a$ and $m=ae^{\l\spo}=a\l x_0^+$ in the usual notation.)
The last term in \qqxxviii\ is the contribution from $\sp
=-\infty$. The function $L(\sm)$ is determined by the
asymptotics of $\e$ as $\sp\to\infty$.
$$
\e=-e^{a_0} d_2 \pmatrix{ e^{{1\over 2}a_0}\cr
-e^{-{1\over 2}a_0}\cr }+
{1\over \sqrt{2}}e^{{3\over 2}a_0} L
\pmatrix{1\cr 0\cr}e^{-{\l\over 2}\sp}
+\Or\left( e^{-\l\sp}\right)\ .
\eqn\qqxx$$
where $a_0=\lim_{\sp\to\infty} \r$. The function $L(\sm)$ is
given explicitly by some integral of functions of $\f$ and
$\r$ [\BILPOS]. In {\it general} we do not have an {\it
explicit}  expression for $\f$, $\r$ at our disposal since
they are given by solving the transcendental relation
between $\f,\r$ and $\O,\x$.  Thus in the generic case one
has to evaluate the integrals numerically. There are
however  certain  cases  where one can still evaluate $L$
analytically  (e.g. as an expansion in $e^{\l\sm}$ as
$\sm\to -\infty$).

Again, we have to fix the functions of integration
$d_1(\sm)$ and $d_2(\sm)$ before we can try to interpret
$M_{\rm B}$ as Bondi-mass. Rather than fixing those
functions ``by hand" and then deducing the properties of
$M_{\rm B}$, we will  impose a few physical requirements
that will actually fix the two functions almost completely.
We require
\pointbegin
for $\k=0$ : $M_{\rm B}=m$. Indeed, in the limit $\k\to 0$,
the model under consideration has no Hawking radiation and
the Bondi-mass should remain constant and equal to the
total energy carried by the in-flux of matter.
\point
as $\sm\to -\infty$ : $M_{\rm B}\to m$. Again, at
$\sm= -\infty$ no Hawking radiation yet had a chance to
occur, and the Bondi-mass must equal $m$.
\point
``causality" : The solution $\e$ in the LDV region shall
not depend on the position and strength of the infalling
matter that is in its causal future.
\par

The last requirement implies $d_1+d_2=0$ as can be seen
from the explicit solution for $\e$ for $\sp<\spo$. The
second reqirement gives $d_2^2(-\infty)=1/2$, while the
first gives $d_2^2(\sm)=1/2+\Or(\k)$. This suggests to take
$d_2^2=1/2$ although it does not exclude a choice like
$d_2^2=1/2+\Or( \k e^{\l\s})$. The latter complication
however seems physically unmotivated and we discard it.
Thus we are lead to take
$$
d_1+d_2=0\ ,\quad d_1^2=d_2^2={1\over 2}\ ,\quad \forall\
\sm\ ,
\eqn\qqxxxii$$
so that in the LDV region $\e=\pm {1\over \sqrt{2}}
\pmatrix{ 1\cr -1\cr}$. It is remarkable that we can
satisfy the above requirements with such a simple choice of
$d_1, d_2$. The conditions \qqxxxii\
 can be written more elegantly as
$$
e^{-2\r}\eb\gf\e\vert_{\sp=+\infty}=-1
\eqn\qqxxxiia$$
which fixes $d_2^2={1\over 2}$, and (note that the l.h.s.
is taken at $\sp=-\infty$, and {\it not} at $\sp=+\infty$)
$$
(\NS \f-\l)\e\vert_{\sp=-\infty}=0
\eqn\qqxxxiib$$
which fixes $d_1+d_2=0$.
With this choice, $\M(\sm,\sp=-\infty)=0$, and
$$
M_{\rm B}(\sm)
={1\over 1-{p\over \l}e^{\l\sm}}
\left[ m +\kk \l
\log\left( 1-{p\over \l}e^{\l\sm}\right)
+\kk p e^{\l\sm}\right]
-\l e^{-\l\sm} L^2(\sm)
\ .
\eqn\qqxxxiic$$

We can evaluate $L(\sm)$ and hence $\mb(\sm)$ for large
negative but finite $\sm$. For the shock-wave scenario
($pe^{\l\spo}=m$),  we arrive at
$$\mb(\sm) \sim m -\k\left( m I_{1/2}
+ {\l\k\over 4}
I_{1/2}^2 \right) e^{-\l\spo} e^{\l\sm} + \Or (e^{2\l\sm})\
, \eqn\qqxxxv$$
where $I_{1/2}$ is a logarithm integral given by
$
I_{1/2} = \int_0^\infty {\rm d} x {e^{-x}\over
x+{\l\k\over 2m}} = -e^{{\l\k\over 2m}}\, {\rm
li}\left(  e^{-{\l\k\over 2m}} \right) >0$.
Let us comment on this equation. First, as already
observed, $\mb$ is constant for $\k=0$: semiclassically there
is no Hawking radiation for $\k=0$. Second, $\mb$ is
decreasing as $\sm$  (i.e. time) increases (at least to the
first order in $e^{\l\sm}$ we computed): Hawking radiation
carries energy away from the black hole.\foot{ Recall that
the we assume $\k\ge 0$.} Note that for
${\l\k\over m} << 1$ the leading term in \qqxxxv\ reads
$\mb \sim m -\k m\log\left( {2m\over \l\k}\right)
e^{-\l\spo} e^{\l\sm}$. This differs from the CGHS
prediction for the very early Hawking radiation by the
extra factor of $\log\left( {2m\over \l\k}\right)$.
However, there is nothing wrong with this difference, since
the RST and CGHS models represent different $\Or (\k)$
corrections to the same classical dilaton-gravity.

Finally we give $\mb(\sm)$ for the
shock-wave scenario at $\sm=\sm_s$, the point where the
singularity and apparent horizon intersect, and the
solution is matched to the LDV. As
shown in ref. [\BILPOS], $\mb(\sm_s)$ can be computed as a
series in ${m\over \l\k}$. It was found that
to first order in ${m\over \l\k}$, we simply have
$$\mb(\sm_s) \sim m +\Or( ({m\over \l\k})^2)\ .
\eqn\qqxxxix$$
Thus if we start with a very small black hole (small $m$)
or a very large number of matter fields (large $\k$), the
black hole is matched to the shifted LDV before any
substantial Hawking radiation has occurred: its mass is
still the initial mass $m$ up to $\Or( ({m\over \l\k})^2)$
corrections. This positive amount of energy must then be
sent off by the thunderpop. In ref. \RST, RST find (up to
their sign ambiguity) that the thunderpop carries energy
${\l\k\over 4}\left( 1-e^{-{4m\over \l\k}}\right)=m +
\Or( ({m\over \l\k})^2)$ in agreement with \qqxxxix.

In conclusion, we have found that our functional $M$ as
given by \tiii\ with $\e$ subject to the differential
equation \tv\ and the boundary conditions \qqxxxiia\ and
\qqxxxiib\ defines a satisfactory Bondi-mass: it is
non-negative, equals the ADM-mass $m$ at $\sm=-\infty$,
decreases for $\k>0$ and is constant for $\k=0$. It also
gives correctly the energy of the thunderpop (at least to
the order we computed), and for $\sm\to -\infty$ has an
expansion in $\k e^{\l\sm}$ as expected.

\chapter{Conclusions}

There is a whole class of exact conformal
2D dilaton
gravity theories
(differing by $\Or(e^{2\f})$ terms)
that all have an action
$S={1\over \pi} \int \left[
\k\dpl\O\dm\O -\k\dpl\x\dm\x +\l^2 e^{2(\x-\O)} \right]$.
The corresponding \eoms are exactly solvable. There is
{\it no} Jacobian in a $\O,\x$ functional integral
formulation.

These theories have (semiclassical) sypersymmetric
extensions. If one insistes on exact superconformal
invariance, the bosonic part is not exactly conformal, and
vice versa.

The Regge-Teitelboim method gives a well-defined (constant)
total ADM-energy (true energy).

We can prove a positive energy theorem. The positive energy
however differs from the previous true energy by an
$\Or(\k)$-term. The Bondi energy can be defined to satisfy
all reasonable physical requirements.

\ack

I am grateful to C. Callan and I. Kogan for the very
stimulating collaborations that are summarized in the
larger part of this paper. It is also a pleasure to
acknowledge helpful discussions with Andy Strominger and
Larus Thorlacius.

\refout

\end